\documentclass[aps,prl,twocolumn,groupedaddress,showpacs]{revtex4}

\usepackage{graphicx}
\begin{document}

\title{Pairing in Asymmetrical Fermi Systems with Intra- and Inter-Species Correlations}


\author{Renyuan Liao}
\author{Khandker F. Quader}
\affiliation{Department of Physics, Kent State University, Kent, OH
44242}


\date{\today}

\begin{abstract}

We consider inter- and intra-species pairing interactions in an
asymmetrical Fermi system. Using equation of motion method, we
obtain coupled mean-field equations for superfluid gap
functions and population densities. We construct a phase diagram
across BCS-BEC regimes. Inclusion of intra-species correlations
result in stable polarized superfluid phase on BCS and BCS sides of
unitarity at low polarizations.
For larger polarizations, we find phase separations in BCS and BEC regimes.
A superfluid phase exists for all polarizations deep in BEC regime. Our results
should be apply broadly to ultra-cold fermions, nuclear and quark matter.

\end{abstract}


\maketitle

Pairing in two-species Fermi systems with unequal
population is of great current interest and importance across
a wide range of fields and systems. Examples are
unequal density mixtures of fermionic cold atoms~\cite{Zwierlein06,Partridge06}; arbitrarily
polarized liquid $^3He$; superconductors in external magnetic
field~\cite{Sarma63}, in strong spin-exchange field~\cite{Larkin65,Fulde64,Takada69},
or with overlapping bands~\cite{Suhl59,Kondo63};
isospin asymmetric nuclear matter~\cite{Sedrakian00} and dense quark matter exhibiting
color superconductivity~\cite{Alford01}.
Unequal density cold fermions serve as prototypical systems,
providing an unprecedented window into exploring superfluidity with tunable
repulsive and attractive interactions. These are attained by sweeping across with
s- or p-wave Feshbach resonances,
thereby allowing the study of fermion ground states in
both BCS and BEC regimes.

Among the {\it outstanding} questions in asymmetrical Fermi systems is
the nature of the ground state in the BCS and BEC regimes,
and whether the BCS superfluid state can sustain any finite
imbalance between the species.
Thus, it is important to arrive at a plausible phase diagram
as a function of pairing interaction strength and species imbalance.
Two-species systems are conveniently characterized as two pseudo-spin systems.
It is believed that the BCS ground state in a finite magnetic field
$h$, is robust against spin polarizations for $h~\sim\Delta$,
($\Delta$ being the superconducting gap); beyond this it becomes
unstable to a normal state.
For {\it equal population} cold atom systems, there is
theoretical agreement with experiments that find superfluid states
in both BCS and BEC regimes with a ``smooth crossover''
around the ``unitarity limit''(diverging singlet scattering length $a_s$).

For systems with {\it population imbalance}, various theoretical
scenarios have been proposed~\cite{Bedaque03,Pao06,Sheehy06}.
Mean-field calculations~\cite{Pao06,Sheehy06} find the superfluid
state to be unstable to phase separation into superfluid and normal
states or a mixed phase in the BCS regime; a superfluid state
stabilizes however deep in the BEC regime. Currently there is
intense experimental efforts in unequal density cold fermion atoms.
One experiment~\cite{Partridge06} observed a transition from a
polarized superfluid to phase separation at a polarization $\sim
10\%$ near unitarity on the BEC side.
 \begin{figure}
      \centering
      \includegraphics[scale=0.80]{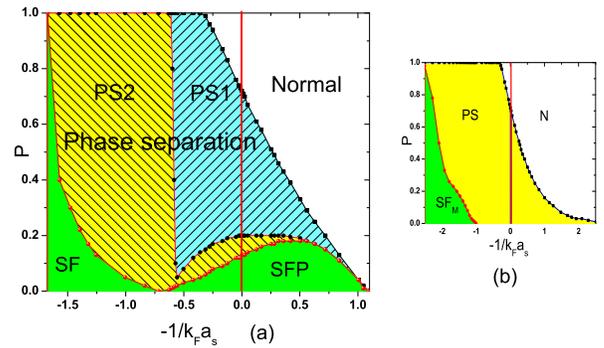}
      \caption{(Color online) Polarization $P$ vs s-wave coupling, $- 1/(k_Fa_s)$ phase diagram
for asymmetrical fermions:(a) with a representative intra-species
correlation strength $g_{1} = 20$,  corresponding to
$1/(k_F^3a_t)$ = 1.25; (b) without intra-species correlations,
$g_1=0$ (also obtained in Ref~\cite{Sheehy06}). Vertical line refers
to the unitarity limit; PS1, PS2, PS to phase separated regions; N
to normal state.}\label{fig:Fig.1}
 \end{figure}
To date, theoretical calculations have mostly considered
{\it inter-species} s-wave interaction, and have ignored
{\it intra-species}  correlations. Ho et al~\cite{Ho06} attempted
to incorporate triplet correlations in a somewhat phenomenological manner;
Huang et al~\cite{Huang06} recently explored the implications for
a FFLO state; Monte Carlo
calculations ~\cite{Carlson05} hint at a  polarized superfluid phase
near unitarity.

{\it In this paper}, we address the issue of
the nature of the zero temperature (T=0) ground state of an asymmetrical Fermi system for arbitrary
repulsive/attractive interaction strength and polarization. We also examine
if the BCS superfluid state can sustain a finite population imbalance.
While the unequal density cold fermion systems may provide a way to test our results,
our paper should have a broader appeal, viz.
electronic superconductivity, nuclear and quark matter superfluidity, etc.
Generally, both inter-species and intra-species correlations may be present in an
asymmetrical Fermi system. These may arise
from the underlying fermionic potentials (atomic, electronic, nucleon-nucleon,
quark-quark) or from effects of the medium, i.e.
``induced'' interactions~\cite{Bulgac06,Quader85}. We include the simplest ones allowed
by symmetry: s-wave contact interaction between the species, and a p-wave interaction
within the species.
Following Leggett~\cite{Leggett80}
and Eagles~\cite{Eagles69}, our discussion is in terms of
BCS-type pairing in both the BCS and BEC regimes, with the
chemical potentials for the two species determined
self-consistently with the pairing gaps.
We do a detailed {\it stability analysis} of the multitude of states obtained
from our equations.

Our findings are {\it dramatically different} from those without intra-species
correlations, Our proposed phase diagram, Fig. 1a, shows that at
T=0, for smaller polarizations, and sufficiently large {\it intra-species} correlations,
gapped polarized superfluidity (hereafter referred to as SFP) becomes {\it stable} on
both BCS and BEC sides
of the ``unitarity'' limit.
Depending on the {\it inter-species} interaction strength, at some polarization, SFP becomes
unstable via a 1st-order transition to {\it phase separation}, denoted by
PS1. PS1 is characterized by a negative ``susceptibility'', $\delta
P/\delta h$; $P$ being the spin-polarization, and $h$ the difference
of the chemical potentials, playing the role of a ``magnetic
field''. For a given {\it intra-species} interaction, and for sufficiently weak
inter-species interaction, SFP and PS1 undergo transitions to the normal
state on the BCS side. The gapped SFP persists into the BEC regime, sustaining progressively
smaller polarizations. Deeper in the BEC regime, we find a superfluid phase (SF)
at all polarizations. In the BEC regime,
in addition to PS1, we find the existence of a somewhat different
phase separated state, PS2, characterized by positive
``susceptibility'', but not satisfying requisite superfluid ground
state stability criteria.

For our detailed study, we consider a two-species Fermi system with unequal
``pseudo-spin'' populations. To allow for both {\it inter-species} and
{\it intra-species}  correlations, and noting that pseudo-spin rotation
invariance would be broken by unequal chemical potentials, we adopt a pairing Hamiltonian given by
\begin{eqnarray}
  H&=&\sum_{k\sigma}\xi_{k\sigma}c_{k\sigma}^{+}c_{k\sigma}\nonumber\\
      &+&\sum_{kk'q\sigma}\frac{g_{kk'}^{\sigma \sigma}}{V} c_{k+q/2\sigma}^{+}c_{-k+q/2\sigma}^{+}c_{-k'+q/2\sigma}c_{k'+q/2\sigma}\nonumber\\
      &+&\sum_{kk'q}\frac{g_{kk'}^{\uparrow\downarrow}}{V} c_{k+q/2\uparrow}^{+}c_{-k+q/2\downarrow}^{+}c_{-k'+q/2\downarrow}c_{k'+q/2\uparrow}
\end{eqnarray}
 where the pseudospin $\sigma=\uparrow,\downarrow$ denote for example the
 two hyperfine states of ultracold Fermi atoms.  $c_{k\sigma}^{+}$ is the fermion
 creation operator with kinetic energy  $\xi_{k\sigma}=\epsilon_{k\sigma}-\mu_{\sigma}$; $\mu_{\sigma}$
is the chemical potential of each of the species.  $g_{kk'}^{\uparrow\uparrow}$, and
$g_{kk'}^{\downarrow\downarrow}$ are the interactions between the up and down spins respectively,
and $V$ is the volume.
The singlet interaction, $g_{kk'}^{\uparrow\downarrow}$ is taken
to be a constant, $g_o$. This is
usually expressed in terms of s-wave scattering length $a_s$ using
 $(4\pi\hbar^{2}a_s/m)^{-1}=g_o^{-1}+\sum_{k}(2\epsilon_k)^{-1}$.
A mean-field decoupling is attained by introducing three order parameters
or gap functions $(\sigma, \sigma' = \uparrow,\downarrow)$ given by,
 $\Delta_{\sigma\sigma'}(k,q)=-\sum_{k'}g_{kk'}^{\sigma\sigma'}<c_{-k'+q/2\sigma}c_{k'+q/2\sigma'}>$
This results in a  mean-field Hamiltonian given by:
\begin{widetext}
\begin{eqnarray}
 H^{MF}&=&\sum_{k\sigma}\xi_{k\sigma}c_{k\sigma}^{+}c_{k\sigma}
           -\sum_{k,\sigma}\Delta_{\sigma \sigma}(k,q)c_{k+q/2\sigma}^{+}c_{-k+q/2\sigma}^{+}
           -\sum_{k,\sigma}\Delta_{\sigma \sigma}^{*}(k,q)c_{-k+q/2\sigma}c_{k+q/2\sigma}
           -\sum_{k.\sigma} |\Delta_{\sigma \sigma}(k,q)|^{2}/g_{kk}^{\sigma \sigma}\nonumber\\
           &-&\sum_{k}\Delta_{\downarrow\uparrow}(k,q)c_{k+q/2\uparrow}^{+}c_{-k+q/2\downarrow}^{+}
           -\sum_{k} \Delta_{\downarrow\uparrow}^{*}(k,q)c_{-k+q/2\downarrow}c_{k+q/2\uparrow}
           -\sum_{k} |\Delta_{\downarrow\uparrow}(k,q)|^{2}/g_{kk}^{\downarrow\uparrow}
\end{eqnarray}
\end{widetext}

We employ the {\it equation of motion} method using imaginary time normal
and anomalous Matsubara Green's functions, $G_{\sigma
\sigma'}(k,\tau)$,  $F_{\sigma \sigma'}(k,\tau)$, respectively, and
our mean-field Hamiltonian, $H^{MF}$.
The coupled equations in terms of $\Delta_{\sigma \sigma'}$ are given by
$G_{\sigma \sigma'}(k,\tau)$  and $F_{\sigma \sigma'}(k,\tau)$:
\begin{eqnarray}
    \partial_{\tau}G_{\sigma\sigma'}(k,\tau)&=&-\delta(\tau)\delta_{\sigma\sigma'}-\xi_{k+q/2\sigma}G_{\sigma\sigma'}(k,\tau)\nonumber\\
    & &+\sum_{\sigma''}\Delta_{\sigma''\sigma}(k,q)F_{\sigma''\sigma'}(k,\tau)\\
    \partial_{\tau}F_{\sigma\sigma'}(k,\tau)&=&\xi_{-k+q/2\sigma}F_{\sigma\sigma}(k,\tau)\nonumber\\
    & &+\sum_{\sigma''}\Delta_{\sigma\sigma''}^{*}(k,q)G_{\sigma''\sigma'}(k,\tau)
\end{eqnarray}
where $\tau$ is the imaginary time variable. These equations may be Fourier transformed in the usual way with
$\tau\rightarrow iw_n$, $\partial_{\tau}\rightarrow -iw_n$, where $iw_{n}=(2n+1)\pi/\beta$
are the Matsubara frequencies, $n$ being an integer and $\beta=1/k_BT$.

Here we focus on a superfluid condensate of pairs with {\it zero center-of-mass momentum}, $q$.
Thus, we do not consider the $q
\ne 0$ Fulde-Ferrel-Larkin-Ovchinnikov(FFLO)
state~\cite{Fulde64,Larkin65}, but which may also be studied within
this scheme. Solving the Fourier transformed equations at $q=0$, we obtain the
{\it 2-point correlation functions}:
\begin{eqnarray}
  G_{\sigma\sigma'}(k,iw_n)&=&\frac{\delta_{\sigma\sigma'}f_{\sigma}+\delta_{\sigma-\sigma'}(\Delta\Delta_\uparrow
  b_\downarrow-\Delta\Delta_\downarrow\Delta_\uparrow)}{D}\nonumber\\
  F_{\sigma\sigma'}(k,iw_n)&=&\frac{\delta_{\sigma\sigma'}f_{\sigma\sigma}+\delta_{\sigma-\sigma'}f_{\sigma-\sigma}}{D}
\end{eqnarray}
where
$f_{\sigma}=\Delta^2b_{-\sigma}+\Delta_{-\sigma}^2b_{\sigma}-a_{-\sigma}b_{\sigma}b_{-\sigma}$;
$f_{\sigma\sigma}=\Delta^2\Delta_{-\sigma}+\Delta_{\sigma}\Delta_{-\sigma}^2-\Delta_{\sigma}a_{-\sigma}b_{-\sigma}$;
$f_{\sigma-\sigma}=\Delta(\Delta^2+\Delta_{\sigma}\Delta_{-\sigma}-a_{\sigma}b_{-\sigma})(\delta_{\sigma\downarrow}-\delta_{\sigma\uparrow})$;
$D=(\Delta^2+\Delta_{\uparrow}\Delta_{\downarrow})^2+a_{\uparrow}a_{\downarrow}b_{\uparrow}b_{\downarrow}
   -\sum_{\sigma}(\Delta^2a_{\sigma}+\Delta_{\sigma}^2a_{-\sigma})b_{-\sigma}$; with
$a_{\sigma}=\xi_{k+q/2\sigma}-iw_n$, $b_{\sigma}=-\xi_{-k+q/2\sigma}-iw_n$.
We have set $\Delta_{\uparrow\uparrow} \equiv \Delta_{\uparrow}$;
$\Delta_{\downarrow\downarrow} \equiv \Delta_{\downarrow}$;
$\Delta_{\uparrow\downarrow} \equiv \Delta$.  The {\it excitation
spectrum} can be found by examining the poles of the Green's
functions, yielding the quasiparticle energies,
\begin{equation}
   E_{k\pm}^2=(iw_n)^2 = (\alpha\pm\sqrt{\beta})/2
\end{equation}
where
$\alpha=\xi_{k\uparrow}^2+\xi_{k\downarrow}^2+2\Delta^2+\Delta_1^2+\Delta_2^2$, and
$\beta=\left[(\xi_{k\uparrow}^2-\xi_{k\downarrow}^2)+(\Delta_1^2-\Delta_2^2)\right]^{2}+4\Delta^2\left[(\xi_{k\uparrow}-\xi_{k\downarrow})^2+(\Delta_1-\Delta_2)^2\right]$.
Various quantities can now be obtained from our 2-point correlation functions.
Thus, particle concentrations, $n_{\sigma}$  for the two species ($\sigma = \uparrow,\downarrow$) are given by
\begin{eqnarray}
&n_{\sigma}&=\sum_{k}\left<c_{k\sigma}^+c_{k\sigma}\right>
            =\sum_{k}\sum_{iw_n}\frac{1}{\beta}G_{\sigma\sigma}(k,iw_n)e^{iw_n0^+}\nonumber\\
             &=&\sum_{k}\sum_{l=\pm}(-1)^{\lambda}\left[\frac{n_F(E_{kl})f_{\sigma}(k,E_{kl})
                    - n_F(-E_{kl})f_{\sigma}(k,-E_{kl})}{2E_{kl} (E_{k+}^2-E_{k-}^2)}\right]\nonumber\\
\end{eqnarray}
where $\lambda$ is even for $l=+$, and odd for $l=-$; $n_F(E_{kl})$
are the Fermi functions. Likewise the three gaps equations are given
by $(\sigma,\sigma' = \uparrow,\downarrow)$:
\begin{widetext}
\begin{eqnarray}
\Delta_{\sigma\sigma'}
=-\sum_{k}g_{kk}^{\sigma\sigma'}\sum_{iw_n}\frac{1}{\beta}F_{\sigma\sigma'}^{*}(k,iw_n)e^{iw_n0^+}
=-\sum_{k}\sum_{l=\pm}(-1)^{\lambda}g_{kk}^{\sigma\sigma'}\left[\frac{n_F(E_{kl})f_{\sigma\sigma'}(k,E_{kl}) - n_F(-E_{kl})f_{\sigma\sigma'}(k,-E_{kl})}{2E_{kl}
(E_{k+}^2-E_{k-}^2)}\right]
\end{eqnarray}
\end{widetext}
The above five equations are {\it coupled}, and can be solved
self-consistently for the three gap functions, $\Delta, \Delta_{\uparrow}, \Delta_{\downarrow}$
for either fixed particle concentrations, $n_{\uparrow}$, $n_{\downarrow}$, or fixed
chemical potentials, $\mu_{\uparrow}$, $\mu_{\downarrow}$.

We assume {\it equal masses} for the two species, and take the particle
spectrum to be $\epsilon_k=\hbar^2k^2/2m$.
We adopt standard definitions: polarization,
$P=(n_1-n_2)/(n_1+n_2)$; mean chemical potential
$\mu=(\mu_1+\mu_2)/2$; chemical potential difference
$h=(\mu_1-\mu_2)/2$; Fermi momentum $k_{F\sigma}
=(6\pi^2n_{\sigma})^{1/3}$. Since
$\sum_kf(k)\rightarrow\int_{0}^{\infty}
f(k)\frac{d^3k}{(2\pi)^3}\rightarrow\int_{0}^{\infty}
k_F^3f(k/k_F)\frac{d^3(k/k_F)}{(2\pi)^3}$, we can scale quantities
having dimension of energy to $\epsilon_F$.
The {\it inter-species} interaction $g_o$ is expressed in terms of
coupling constant $\eta=-1/(k_{F}a_s)$.
For the {\it intra-species} triplet interaction, we take the separable form
$g_{kk'}^{\sigma\sigma}=g_1\omega(k)\omega(k')Y_{10}(\hat{k})Y_{10}(\hat{k'})$,
where we have taken $g_{\uparrow\uparrow} = g_{\downarrow\downarrow} \equiv \tilde{g_1}$.
More generally the $m=\pm1$ terms would also be present; however
this choice allows us to explore the consequences of intra-species
correlations while keeping the calculations tractable. As a check,
we also consider different types of momentum dependence: (i)
$\omega(k)\propto const$; (ii) $\omega(k)\propto k_0k/(k_0^2+k^2)$,
a generalization of the Nozieres and Schmitt-Rink
scheme~\cite{NSR85}; (iii) $\omega(k)\propto \exp{[-(k/k_0)^2]}$, a
Gaussian interaction; $k_o$ being a cut-off momentum; these give
qualitatively similar behavior. The first two forms of interaction
require regularization due to ultraviolet divergence, while the
third does not. With regularization, $g_1$ can be expressed in terms
of a triplet scattering volume $a_t$~\cite{Iskin06}:
$(4\pi\hbar^{2}k_o^2a_t/m)^{-1}=g_1^{-1}+\sum_{k} w(k)^2/(2
\epsilon_k)$. Thus, $(3 n/2 \epsilon_F) \tilde{g_1} \equiv g_1$ in
our plots can be easily expressed in terms of $a_t$; e.g. $g_1 = 20
$ corresponds to $1/(k_F^3a_t) = 1.25$.

For arbitrary inter-species s-wave and and intra-species p-wave pairing interactions,
and population imbalances, we obtain self-consistent solutions of the
$T=0$ gap functions, $\Delta, \Delta_{\uparrow},
\Delta{\downarrow}$, and chemical potentials, $\mu_{\sigma}$. On the
BCS side, for a given $g_o$, the
$\uparrow \downarrow$ gap $\Delta$ {\it decreases} with increasing
intra-species interaction strength $g_1$, while at the same time
both $\Delta_{\uparrow}, \Delta_{\downarrow}$ ($\Delta_{\uparrow} \ne \Delta_{\downarrow}$)
increases, crossing at some value of $g_1$. The suppression of $\Delta$ is more pronounced
at larger polarizations.

A proper construction of the asymmetrical Fermi system {\it phase diagram}
requires a determination of stable ground states out of the manifold of paired
condensates given by our equations~\cite{Sheehy06,Viverit00}. Accordingly, we
carefully consider the stability criteria. The
mean-field {\it ground state energy} as a function of the gaps
at different polarizations, P is given by:
 \begin{eqnarray}
&E_G&(\Delta,\Delta_{\uparrow}.\Delta_{\downarrow})=\left<\Psi|H^{MF}|\Psi\right>\nonumber\\
&=&E_o + \sum_{k\sigma}\left[\xi_{k\sigma}G_{\sigma\sigma}(k,\tau=0^-)-2\Delta_{\sigma}F_{\sigma\sigma}(k,\tau=0^-)\right]\nonumber\\
&-&\sum_{k}2\Delta F_{\downarrow\uparrow}(k,\tau=0^-)
 \end{eqnarray}
where
$ E_o=-|\Delta_{\uparrow}|^2/g_1 - |\Delta_{\downarrow}|^2/g_1 - |\Delta|^2/g_o$.
To find the stability of the polarized superfluid state SFP,
we construct the 3x3 stability matrix out of all partial derivatives
$\frac{\partial^2 E_G}{\partial \Delta_i \partial \Delta_j}$ ($\Delta_{i,j} = \Delta, \Delta_{\uparrow},
\Delta_{\downarrow}$), and check for positive definiteness of the determinant of the
matrix, and of all its upper-left sub-matrices.
We supplement this with analysis of
the ``susceptibility'' $\partial P/\partial h$. Thus, for a given $g_1$, the
stable polarized superfluid state SFP, in both BCS and BEC regimes,
is characterized by $E_G$ with a {\it global minimum} at non-zero gaps
and self-consistently determined values of
$\mu_{\sigma}$'s, and
$\frac{\partial P}{\partial h} > 0$.
SFP sustains larger polarizations for progressively larger $g_1$.

Similar to the case without intra-species correlations, $g_1$, there
exists a maximum polarization, $P_{max}$ on the BCS side, beyond which
we find no solution to the coupled equations;  this determines the SFP/PS1
$\leftrightarrow$ N boundary (Fig. 1a). $P_{max}$ is slightly decreased at unitarity by
$g_1$. It decreases with increasing $\eta= - 1/k_Fa_s$. For a fixed $g_1$,
close to both BCS and BEC sides of unitarity, and for small polarizations,
unlike-spin pairing has appreciable value in the polarized superfluid SFP.
However away from unitarity on the BCS side,
$\Delta_{\uparrow \downarrow}$ decreases, and $\uparrow \uparrow$ and $\downarrow \downarrow$
pairing becomes more dominant in SFP, as inter-species interaction becomes relatively
weak compared to intra-species interaction.
On the BEC side away from unitarity, on the other hand,
$\Delta_{\uparrow \downarrow}$ is more dominant, and with $\Delta_{\uparrow}$
and $\Delta_{\downarrow}$ becoming negligible, SFP becomes unstable to phase separation, PS2.
A SF phase emerges deep in the BCS regime with predominantly unlike-spin pairing at
low polarizations and majority spin pairing at higher polarizations.
\begin{figure}
     \centering
     \includegraphics[scale=0.70]{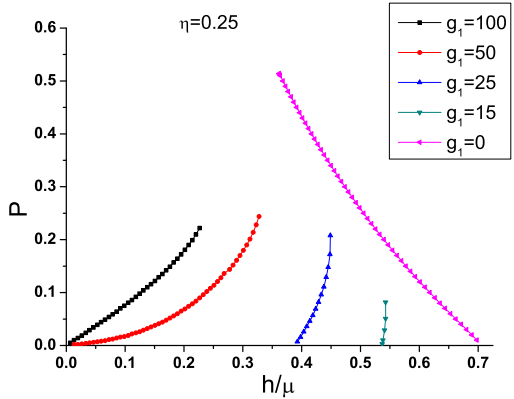}
     \caption{(Color online) Polarization $P=\frac{n_{\uparrow} - n_{\downarrow}}{n_{\uparrow} + n_{\downarrow}}$ vs
$h/\mu = \frac{\mu_{\uparrow} - \mu_{\downarrow}}{\mu_{\uparrow} + \mu{\downarrow}}$ ($h$ acts like
a ``magnetic field'') for different intra-species strengths $g_1$ at a fixed
inter-species coupling $\eta = 0.25$.}\label{fig:Fig 2}
     \end{figure}

The region PS1 in Fig 1a is characterized by {\it negative}
``susceptibility'', $\frac{\partial P}{\partial h}$ and
doe not satisfy the stability matrix criteria.
For a given $P$ and
$h$, $E_G$ is a {\it maximum} at the non-zero gap solutions,
separating two local minima -- a feature of phase separation into a
normal and a superfluid  component by
1st-order phase transition. In this context, it is instructive to
study $P$ as a function of $h/\mu$  for different values of $g_1$,
for a fixed $\eta$ (shown in Fig.2 for
$\eta=0.25$ (BCS side)). For $g_1 = g_1^c \approx 15$, the slope is
vertical  ($g_1^c$ is the value of $g_1$ at which maximum
polarization, $P_{max}$ occurs for a given $\eta$).
For $g_1 \le g_1^c$, the slope is {\it negative}, corresponding to
the BCS superfluid state being unstable to the normal state for $h >
\Delta$, but robust against polarization for $h < \Delta$. For $g_1
> g_1^c$, the singlet superfluid state can sustain a {\it finite
polarization}, which exhibits a behavior over the range given by: $P
\propto ah + bh^3 + c$; a,b,c being constants. The linear behavior
is achieved for larger values of $g_1$, and at low polarizations. In
examining $P$ vs $h/\mu$ behavior {\it beyond} $P_{max}$, we  find,
for a given $g_1$, {\it two} solutions of $P$ corresponding to one
value of $h/\mu$. To make a connection to Fig. 1a, obtained for $g_1
= 20$, we note that the allowed range of polarizations for SFP in
$P$ vs $h/\mu$ considerations corresponds to the $\eta=0.25$
vertical line, terminating at the SFP-PS1 boundary (Fig 1a). The
same line extended from SFP-PS1 boundary to PS1-N boundary
correspond to the polarization range bounded by the two solutions of
$P$ at a $h/\mu$ in Fig.2. The region PS2 in BEC regime, though
characterized by $\frac{\partial P}{\partial h} > 0$, is not a
stable superfluid phase, since the stability matrix condition cannot
be satisfied. The line separating PS1 and PS2 is probably a
metastable line, the position of  which depends on the p-wave
interaction strength.

In summary, we find that the inclusion of intra-species correlations
in asymmetrical Fermi systems results in a stable polarized
superfluid phase SFP at low polarizations on both BCS and BEC sides
of unitarity. We have discussed the nature of the paired states and
transition to phase separated states. The SF phase obtained in the
deep BEC regime in the case without intra-species correlations, also
emerges here with dominant unlike-species pairing, accompanied by
weaker majority-species pairing. Our results should be of broad
interest as it should be of relevance to any asymmetrical Fermi
system, with proper choice of interaction parameters. Here, our
choice of parameters appear to agree with cold atom
experiment~\cite{Partridge06} that found a SF to PS transition
around $\sim 10\%$ polarization around unitarity on the BEC side.
Also, the maximum polarization $\sim 70\%$ at unitarity is in
agreement with experiments. Further experiments at low polarizations
on both sides of unitarity are needed to test our detailed results.
Experiments that measure differences in momentum distributions of
two species could provide further test. Finally, our phase diagram
indicates a tricritical point (SFP,PS1,N phases) at low polarization
on the BCS side, in addition to one on the BEC side at $P \sim 1$.
This should lead to interesting study of the evolution of these two
tricritical points at finite-T; we are exploring these effects.

We would like to thank D. Allender, K. Bedell, J. Engelbrecht, S. Gaudio, R. Hulet, and M. Widom
for fruitful comments and discussions. We also acknowledge support from ICAM.

\end{document}